\documentclass[letterpaper]{article} % DO NOT CHANGE THIS
\usepackage{aaai25}  % Change 'submission' to 'final'
\usepackage{times}  % DO NOT CHANGE THIS
\usepackage{helvet}  % DO NOT CHANGE THIS
\usepackage{courier}  % DO NOT CHANGE THIS
\usepackage[hyphens]{url}  % DO NOT CHANGE THIS
\usepackage{array}
\usepackage{graphicx} % DO NOT CHANGE THIS
\usepackage{natbib}  % DO NOT CHANGE THIS AND DO NOT ADD ANY OPTIONS TO IT
\usepackage{caption}
\usepackage{booktabs}% DO NOT CHANGE THIS AND DO NOT ADD ANY OPTIONS TO IT
\usepackage{algorithm}
\usepackage{algorithmic}

\usepackage{newfloat}
\usepackage{listings}
\DeclareCaptionStyle{ruled}{labelfont=normalfont,labelsep=colon,strut=off} 
\floatstyle{ruled}
\newfloat{listing}{tb}{lst}{}
\floatname{listing}{Listing}

\title{Towards Interactive Evaluations for Interaction Harms in Human-AI Systems}

\author{
    Lujain Ibrahim\textsuperscript{\rm 1}, 
    Saffron Huang\textsuperscript{\rm 2}, 
    Lama Ahmad\textsuperscript{\rm 4}, 
    Umang Bhatt\textsuperscript{\rm 3}, \\
    Markus Anderljung\textsuperscript{\rm 5}
}
\affiliations{
    \textsuperscript{\rm 1}University of Oxford\\
    \textsuperscript{\rm 2}Collective Intelligence Project\\
    \textsuperscript{\rm 3}NYU Center for Data Science\\
    \textsuperscript{\rm 4}OpenAI\\
    \textsuperscript{\rm 5}Centre for the Governance of AI
}

\begin{document}

\maketitle

\begin{abstract}
Current AI evaluation methods, which rely on static, model-only tests, fail to account for harms that emerge through sustained human-AI interaction. As AI systems proliferate and are increasingly integrated into real-world applications, this disconnect between evaluation approaches and actual usage becomes more significant. In this paper, we propose a shift towards evaluation based on \textit{interactional ethics}, which focuses on \textit{interaction harms}—issues like inappropriate parasocial relationships, social manipulation, and cognitive overreliance that develop over time through repeated interaction, rather than through isolated outputs. First, we discuss the limitations of current evaluation methods, which (1) are static, (2) assume a universal user experience, and (3) have limited construct validity. Drawing on research from human-computer interaction, natural language processing, and the social sciences, we present practical principles for designing interactive evaluations. These include ecologically valid interaction scenarios, human impact metrics, and diverse human participation approaches. Finally, we explore implementation challenges and open research questions for researchers, practitioners, and regulators aiming to integrate interactive evaluations into AI governance frameworks. This work lays the groundwork for developing more effective evaluation methods that better capture the complex dynamics between humans and AI systems.\end{abstract}

\section{Introduction}
Artificial intelligence (AI) model evaluations, broadly defined as systematic empirical assessments of AI models' capabilities and behaviors, have become central to developers' and regulators' efforts to ensure that AI systems are sufficiently safe. Governments have emphasized the importance of model evaluations for risks ranging from discrimination to cybersecurity, establishing dedicated bodies such as the UK's AI Security Institute to carry them out \citep{UK_Government_2024}; AI labs, including OpenAI with its Preparedness Framework and Anthropic with its Responsible Scaling Policy \citep{OpenAI_2023, Anthropic_2023}, propose using model evaluations to monitor and mitigate misuse and catastrophic risks; and academic researchers are developing evaluation datasets at unprecedented rates \citep{rottger2024safetyprompts}.

The growing importance of model evaluations has been accompanied by increased scrutiny, with researchers identifying both the unique challenges of evaluating generative AI systems compared to traditional machine learning systems, and broader concerns about validity, replicability, and quality \citep{weidinger2023sociotechnical, liao2023rethinking,raji2022fallacy,wallach2025position}. A prominent thread in these concerns is the disconnect between current evaluation approaches and real-world use of AI systems---mostly in the form of large language models (LLMs)---today. While this is rapidly changing, evaluations have historically relied primarily on static, isolated tests that assess models based on their responses to individual prompts \citep{chang2024survey, weidinger2023sociotechnical}. Yet this approach is at odds with real-world AI use, where applications increasingly depend on sustained back-and-forth interaction---from AI friends and companions engaging millions of users in daily conversations to agentic AI systems that take actions on users' behalf.

This mismatch between evaluation methods and real-world use has become more consequential as AI systems proliferate in homes, schools, and workplaces. Many of the public's, developers', and policymakers' most pressing concerns---such as inappropriate human-AI relationships, social influence and manipulation, and overreliance---stem from patterns of repeated interaction, not isolated model outputs. While once hypothetical concerns, there are now many documented cases of these harms. For example, sustained use of AI companions has been linked to serious impacts on vulnerable users, including cases now the subject of litigation against major developers, a Federal Trade Commission inquiry into companion chatbots, and a U.S. Senate Judiciary hearing \citep{ftc_6b_companion_2025, senate_judiciary_2025}. In such cases, the harms accumulate gradually, as a user who relies on an AI companion for daily conversation over weeks or months may experience changes in decision-making and emotional state even when each individual response passes standard safety evaluations. This type of harm, which emerges through sustained interaction, is missed by conventional evaluation methods \citep{alberts2024should}.

In this paper, we argue that while traditional machine learning evaluation paradigms are useful for many purposes, they fail to fully capture harms that arise from extended human interaction with AI systems. We propose a shift towards evaluating AI systems through the lens of \textit{interactional ethics}, which focuses on the interaction harms that emerge through sustained engagement with AI as social and relational actors \citep{alberts2024should}. \footnote{Our examples focus on LLMs and text interactions because they provide the richest available data for studying sustained human-AI engagement patterns. However, the principles discussed could be applied to multi-modal systems as they mature.} Human participation is already playing a growing role in AI research, with studies involving human data making up approximately 9\% of papers at top computer science venues AAAI and NeurIPS from 2021 to 2023 \citep{mckee2024human}. Concurrently, social science research is increasingly conducting large-scale experiments with AI systems to understand their impact on human behavior \citep{costello_pennycook_rand_2024}. Our approach unites these complementary insights from human-computer interaction (HCI), natural language processing, and the social sciences.

We begin by analyzing how current machine learning evaluation approaches fall short in assessing interaction harms due to their (1) static nature, (2) assumption of a universal user experience, and (3) limited construct validity. Drawing on decades of HCI and social science research, we then propose organizing principles for evaluating generative models, structured around designing interaction scenarios, measuring human impact, and determining appropriate forms of human participation. We conclude by outlining key challenges and opportunities for researchers, companies, and regulators seeking to implement interactive evaluations of generative AI systems, including considerations of scalability, standardization, and practical integration with existing frameworks.

\section{An Overview of the Generative AI Evaluation Landscape}

We begin by examining contemporary approaches to ethics and safety evaluations of generative AI systems, their methodologies, primary focus areas, and limitations.

\subsection{The Current State of AI Safety Evaluations}
We follow existing work in adopting a wide definition for “safety” which encompasses model behaviors associated with various taxonomized harms \citep{weidinger2022taxonomy, shelby2023sociotechnical}. Examples of these include different types of biases \citep{parrish2021bbq}, toxicity \citep{hartvigsen2022toxigen}, and “dangerous capabilities” like persuasion and cybersecurity risks \citep{phuong2024evaluating, li2024wmdp}. Safety evaluations of generative AI systems build on a rich history of ethical considerations in the field of natural language processing (NLP), prior to the advent of large pre-trained models, where researchers have long grappled with issues of harm from language technologies \citep{dev2021measures}.

Recent reviews of existing safety evaluations show that the majority are focused on evaluating individual model responses to prompts curated by researchers targeting various model behaviors \citep{rottger2024safetyprompts, weidinger2023sociotechnical}. The evaluations that do involve human subjects or span multiple dialogue turns tend to take the form of “red teaming campaigns” or automated adversarial testing that assumes malicious user intent \citep{Zhang_2024, perez2022red}. These reviews point to two key gaps: (1) a \textit{methodological} gap: multi-turn evaluations remain scarce and, where they exist, are framed around adversarial rather than everyday use, and as a result, (2) a \textit{coverage} gap: since interaction harms require evaluations that go beyond assessing model behavior in isolated, single-turn interactions.

A small but growing body of work has begun addressing these gaps, by conducting studies that utilize production chat logs \citep{phang2025investigating} or user simulations \citep{ibrahim2025multi, zhou2024haicosystem} to investigate risks related to affective use of AI systems. Our work builds on and extends these emerging approaches to interactive evaluation.

\subsection{Critiques of Current Evaluations}
With the growing adoption of model evaluations, research has increasingly pointed out validity issues in how they assess generative models. Some researchers have questioned the external validity of current evaluations, noting that benchmark tasks poorly mirror real-world use cases and may not capture how models actually behave outside controlled settings \citep{ouyang2023shifted, liao2023rethinking}. Others have shown that current evaluations lack sufficient construct validity, especially when operationalizing complex concepts like fairness \citep{raji2021ai,blodgett2021stereotyping}. 

In broader methodological reflections, researchers have also argued that LLMs should not be evaluated using frameworks designed for assessing humans, since LLMs exhibit distinct sensitivities, for example to prompt variations \citep{mccoy2023embers}. These challenges have led researchers to advocate for an interdisciplinary approach that draws on diverse traditions: the social sciences' emphasis on measurement validity \citep{wallach2025position}, HCI's focus on bridging technical capabilities and social requirements \citep{liao2023rethinking}, and cognitive science's frameworks for analyzing system objectives \citep{mccoy2023embers}.

\subsection{Methods for Studying Human-Computer Interaction}
Research on human--computer interaction (HCI) and the broader social and behavioral sciences offers diverse ways to involve humans in studying computational systems --- from having humans as research subjects who actively engage with systems to having them serve as external observers and annotators. Building on this work, we distinguish between static and interactive evaluation approaches by whether they capture either the multi-turn dynamics of an interaction or its impact on users over time \citep{lee2022evaluating}.
Static evaluations focus on isolated inputs and outputs, without considering the progression of the conversation. Interactive evaluations, in contrast, track multi-turn exchanges and how model behavior evolves and adapts to a responsive user, and/or measure how interacting with a model affects users over time. This distinction is not about who performs the evaluation: automated tests that simulate realistic, adaptive user behavior can be interactive when they capture these dynamics, whereas human studies that examine only single-turn responses remain static. In practice, this is a spectrum rather than a strict binary. Interactive evaluations can be \emph{experimental}---conducted in controlled, often lab-like or randomized settings to isolate specific variables---or \emph{observational}, analyzing existing multi-turn interactions from production chat logs to surface patterns and correlations between model behaviors and reported user experiences \citep{hariton2018randomised, 2003xiii}.

\section{Why Current Evaluations Approaches are Insufficient for Assessing Interaction Harms}
\textbf{First, most current evaluations are single-turn, measuring model behavior on isolated inputs and outputs. This does not capture real-world use, where people interact with models over many dialogue turns rather than in one-off exchanges.
}

Unlike many problematic model outputs — such as toxic language or factual errors, which can be judged from a single response — the behaviors that drive interaction harms unfold through the back-and-forth of a conversation, and are therefore difficult to detect with single-turn evaluation.

Two features of multi-turn interaction make these harms invisible to single-turn evaluation. First, harmful behavior may be \emph{compositional}, where each individual response can appear benign, but the same pattern repeated across a conversation produces a concerning cumulative effect. Second, harmful behavior may be \emph{emergent}, surfacing only after several turns rather than in an opening response. In both cases, an evaluation that examines a single prompt--response pair will systematically miss the behavior of interest, which is why assessing interaction harms demands multi-turn evaluation that tracks how model behavior develops over a conversation. Recent work shows, for example, that anthropomorphic behaviors---a model's expressions of selfhood, emotion, or relationship---surface disproportionately as a conversation unfolds rather than in the opening response \citep{ibrahim2025multi}, and so remain largely invisible to single-turn evaluation.

\textbf{Second, current evaluations collapse the diversity of user groups into a single, “universal” user. This overlooks how various demographic groups engage with models uniquely, and how models may tailor their responses to distinct user populations.
}
Current evaluation approaches rely heavily on standardized datasets of prompts written by researchers or crowdworkers. Although controlled test sets are a necessary starting point for systematic evaluation, they implicitly assume a uniform user. In reality, users vary by demographics, domain expertise, technical knowledge, and psychological state, and engage with systems in correspondingly different ways \citep{10.1145/3531146.3533182, ibrahim2024characterizing}.

Beyond explicit personalization, models increasingly engage in implicit user modeling, forming internal representations that shape their responses \citep{chen2024designing,staab2023beyond}. As a result, they adapt their behavior to perceived user attributes \citep{viegas2023system}. Some adaptations seem benign, such as mirroring a user's vocabulary as the conversation progresses, but many others are not. Models have been shown to vary their refusals of dangerous queries by perceived user identity or persona \citep{ghandeharioun2025s}, and to respond more toxically to African American English than to Standard American English \citep{hofmann2024ai}. Accounting for this variance is especially important for generative models, whose adaptive, probabilistic behavior introduces further uncertainty \citep{jin2024implicit}. It is not just demographic variables that may impact how a model responds. The emotional state a user expresses, the vulnerability they might disclose, and the relational stance they adopt towards the chatbot (e.g., treating it like a friend vs a tool) may also shape how a model responds.

\textbf{Third, current evaluations are focused on model outputs rather than their impact: while they can measure what a model \textit{does}, they cannot measure its \textit{impact} on human perception and behavior. This is a construct-validity gap for interaction harms, and a different set of metrics is needed to assess that impact directly.}

Static evaluation methods can effectively identify harmful content, such as toxic or illegal outputs \citep{wallach2025position}. However, they cannot address interaction harms, where the harm lies not in the content itself but in its effect on users. Unlike identifying illegal content, where the task is complete once the content is classified against established criteria, interaction harms require establishing causal links between model behaviors and measurable human outcomes. Detecting that an LLM response contains manipulative language, for instance, does not tell us whether this language successfully influenced the user's decision, as that is about the user, not the text. Similarly, detecting a racially-biased LLM does not tell us whether it successfully shifts a user's decisions in high-stakes settings. And, classifying an LLM response as friendly or empathetic does not reveal whether or not it fostered unhealthy emotional dependence in a user \citep{phang2025investigating}. In each case the model output can be classified on its own, but its impact can be established only by measuring the human.

Thus, establishing construct validity for interaction harms requires quantifying specific human impacts, such as shifts in user beliefs, decision-making patterns, affective states, and dependency levels. Importantly, recent work shows that once these links between model behaviors and human impacts are validated through human experiments, the identified patterns can be repurposed as efficient static tests---proxies with empirically established predictive validity for real-world impact \citep{ibrahim2025multi}.

\section{Towards Better Evaluations of Interaction Harms}

The growing adoption of AI systems in daily life demands evaluation methods that can capture nuanced human-AI interaction dynamics and their potential harms. Drawing on established methodologies for studying HCI and addressing the limitations identified in the previous section, we propose three organizing principles for developing interactive evaluations. We structure these principles around key challenges: designing ecologically valid scenarios, measuring human impact, and determining appropriate human participation:

\begin{enumerate}
    \item \textbf{Scenario design}: developing ecologically valid contexts that reflect real-world interaction patterns and user objectives
    
    \item \textbf{Impact measurement}: identifying metrics for measuring how model behaviors impact human behaviors
    
    \item \textbf{Participation strategy}: determining appropriate forms of human involvement to balance experimental control and costs with authentic interaction
\end{enumerate}
\subsection{Principle 1: Design Interaction Scenarios Based on User Objectives and Interaction Modes
}

\begin{table*}[h]
    \centering
    \label{tab:use-context}
    \begin{tabular}{>{\raggedright\arraybackslash}p{0.22\textwidth}>{\raggedright\arraybackslash}p{0.22\textwidth}>{\raggedright\arraybackslash}p{0.22\textwidth}>{\raggedright\arraybackslash}p{0.22\textwidth}}
    \toprule
    \textbf{Scenario} & \textbf{Misuse} & \textbf{Unintended harm: personal impact} & \textbf{Unintended harm: external impact} \\
    \midrule
    \textbf{Objective} & User intentionally uses model to inflict harm on another person or group of people & User uses model, gets harmed in the process & User uses model, unintentionally harms another person or group of people \\
    \midrule
  \textbf{Affected parties} & External subjects & User & External subjects \\
  \midrule
    \textbf{Example(s)} & User utilizes model capabilities for planning a biological attack & User overrelies on empathetic model for mental health support, delaying professional help  & User trusts biased model judgment, making a discriminatory hiring decision  \\
    \bottomrule
    \end{tabular}
        \caption{Possible harmful use scenarios and examples of each}

\end{table*}

To systematically evaluate human-AI interaction, we need to consider two key dimensions: why users engage with these systems (their objectives) and how they interact with them (their interaction modes).
In HCI, user goals or objectives have been shown to shape how users engage with systems and thus influence the outcomes of these interactions \citep{subramonyam2024bridging}. Therefore, we provide a categorization of harmful scenarios based on user objectives and affected parties, as shown in Table~\ref{tab:use-context}. These scenarios capture key patterns of harm from current generative AI systems that researchers and practitioners have observed \citep{Mitchell_2024}.

\begin{figure*}[!htbp]
    \centering
    \includegraphics[width=\textwidth]{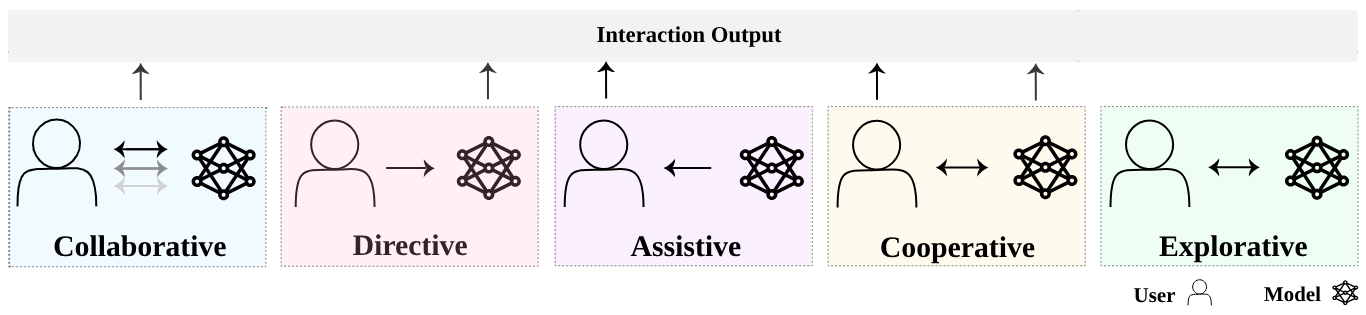} 
    \caption{Taxonomy of human-AI interaction modes}
    \label{fig:widefigure}
\end{figure*}

Beyond their objectives, users’ different patterns of engagement with AI systems shape how their interactions unfold. A single generative model can support multiple interaction modes, from simple question-answering to complex collaborative tasks. Drawing on observed use cases and literature reviews, we identify five prototypical modes of human-model interaction \citep{gao2024taxonomy, ouyang2023shifted, zhao2024wildchat, handler, collins2023evaluating}, visualized in Figure~\ref{fig:widefigure}:
\begin{itemize}
    \item \textbf{Collaborative}: human and model work in tandem towards completing joint goal-oriented tasks (e.g., human and model iteratively write and refine a report together)
    \item \textbf{Directive}: human delegates specific tasks to the model for independent completion (e.g., human instructs model to generate a marketing campaign)
    \item \textbf{Assistive}: model provides supporting input while human maintains primary agency (e.g., model suggests edits while human writes a document)
    \item \textbf{Cooperative}: human and model make distinct contributions toward a shared goal without directly working together (e.g., model generates data visualizations while human writes analysis)
    \item \textbf{Explorative}: human engages in open-ended interaction without specific task goals (e.g., casual conversation or creative brainstorming)
\end{itemize}

\subsection{Principle 2: Identify the Causal Link Between Model Behavior and Human Impact} 
\begin{figure*}[ht]
    \centering
    \includegraphics[width=\textwidth]{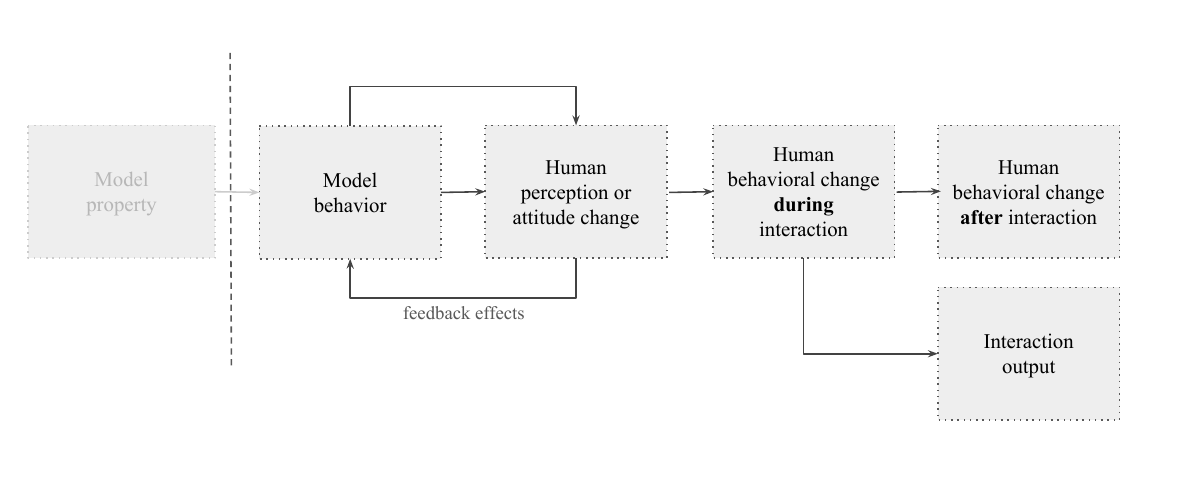} 
    \caption{Example of a causal trace showing how model properties may influence human behavior and human-AI interaction outcomes. Such traces help identify key measurement points for evaluation.}
    \label{fig:figure2}
\end{figure*}

To evaluate interaction harms in a valid way, we must trace how model behaviors influence human users. In Figure~\ref{fig:figure2}, we present an example of such a trace: underlying model properties shape observable model behaviors, which affect users through various pathways — from shaping moment-to-moment interactions to influencing deeper patterns of perception, decision-making, and behavior. Specifying the hypothesized pathways helps determine both what to manipulate in evaluations (model behaviors) and what to measure (human impact). We identify three key measurement targets:
\begin{itemize}
    \item \textbf{Psychological impact}: changes in user perceptions, attitudes and beliefs, and affective states
    \item \textbf{Behavioral impact}: changes in user actions during and after interaction
    \item \textbf{Interaction outputs}: quality and characteristics of the produced interaction artefact
\end{itemize}

These measurement targets can be assessed through both \textit{self-reported} measures — including validated scales for attitudes, perceptions, and affective states — and \textit{behavioral} measures that track concrete actions like response time, task accuracy, and engagement patterns \citep{8404030, CORONADO2022392, 10.1145/3411764.3445423}. In Table~\ref{tab:measurement}, we present a non-exhaustive set of these metrics for assessing both human impact and human-AI performance.

\begin{table*}[ht]
    \centering
    \label{tab:measurement}
    \setlength{\arrayrulewidth}{0.4mm} 
    \renewcommand{\arraystretch}{1.1} 
    \begin{tabular}{p{0.25\textwidth}p{0.33\textwidth}p{0.33\textwidth}}
    \midrule
    \textbf{Evaluation target} & \textbf{Description} & \textbf{Metrics} \\
    \midrule
    Psychological impact & User affective states, or perceptions and attitudes toward the model or the interaction & Usability metrics, user satisfaction surveys, psychometrically validated measures for a specific construct or harm (e.g., Decision Regret Scale \citep{decisionregret}) \\
    \midrule
    Behavioral impact (measured during the interaction) & Observable patterns of behavior recorded during the interaction & Number of queries users made, number of revisions users made, time between queries \\
    \midrule
    Behavioral impact (measured after interaction) & Observable behaviors assessed following the interaction & Adherence to AI suggestions (e.g., choosing to donate when the AI recommends it), disclosing personal information when asked by the AI \\
    \midrule
    Interaction output & Objective quality of output, assessed either automatically (e.g., performance accuracy) or using third-party evaluators (human or LLM) & Consistency of a summary given a document, success rate for completing a task \\
    \bottomrule
    \end{tabular}
        \caption{Metrics to evaluate human-AI interactions}

\end{table*}

\subsection{Principle 3: Structure Human Participation to Balance Validity and Practicality
} Evaluating human-AI interaction can involve varying degrees of human participation: from controlled studies with live participants, to analyses of chat logs between users and deployed systems, to automated user simulations. Each of these different approach is subject to trade-offs in ecological validity, experimental control, cost, and ethical considerations.

Chat logs from real-world interactions provide unparalleled ecological validity and scale. These naturally occurring conversations capture authentic user behaviors and patterns that would be difficult to replicate in controlled settings \citep{tamkin2024clio}. The massive volume of existing logs also enables analysis of rare events and edge cases that controlled studies might miss. Recent work has paired analyses of millions of ChatGPT conversation logs with longitudinal surveys of those users to examine emotional well-being, showing a path forward for using chat logs while also assessing real user impact \citep{phang2025investigating}. However, this observational data is often proprietary and difficult to access, raises serious privacy concerns, and may limit our ability to test specific hypotheses or establish causal relationships \citep{reuel2024open,zhao2024wildchat, ouyang2023shifted}. 

Controlled human-subjects studies are indeed more resource-intensive, but they allow for the systematic examination of specific interaction patterns and psychological outcomes or behavioral changes. However, their artificial nature and high costs make them better suited for targeted investigation of specific hypotheses rather than broad-scale evaluation.

Existing research on user simulations has utilized LLMs to simulate believable human behaviors using a range of architectures from demographic-based models to simulations informed by qualitative interviews \citep{park2024generative,park2023generative}. Recent work has used such simulations to evaluate safety risks, utilizing human experiments to validate the evaluations \citep{ibrahim2025multi, zhou2024haicosystem}. Such simulations can be particularly valuable for exploring scenarios that would be ethically challenging to test with real users. Unlike real human participants, who naturally shape conversations in diverse ways, simulations also allow for controlled trajectories, making it easier to analyze how interactions evolve across different scenarios. However, their limitations in believability, accuracy, and diversity require further study, as recent work has shown that they misportray and flatten representations of marginalized identity groups \citep{wang2025large, agnew2024illusion}. 

\section{Open Challenges and Ways Forward for Interactive Evaluations}

In this paper, we motivate the need to investigate human-AI interaction dynamics that current evaluations miss. However, implementing such evaluations at scale presents several concrete challenges, including ethical questions about studying potentially harmful interactions, practical needs for research infrastructure, and methodological considerations about producing insights that can be actioned by stakeholders. While our design principles outline key considerations for scenarios, measurements, and participation strategies, advancing these methods requires addressing several open questions which we identify below. Concentrated research effort addressing these challenges could significantly advance our ability to evaluate increasingly interactive AI systems.

\subsection{How Can We Ethically Work with Human Participants on Studying Harms? When Are User Simulations Appropriate Replacements?} Studying interaction harms poses inherent ethical challenges: we need to understand potentially harmful dynamics without exposing human participants to those same harms. This requires careful consideration of when and how to involve human participants. For participant protection, AI researchers should adopt existing practices from fields experienced in providing adequate participant training, thoughtful debriefing, and active feedback collection to mitigate experimental harms \citep{mckee2024human}. Additionally, simulating human interactions using LLM-based methods offers a promising direction for studying scenarios where minimizing harm is challenging, such as working with vulnerable populations (e.g., young people's interactions with AI systems or mental health risks from AI systems). Recent work suggests these simulations can believably represent human behavior \citep{park2023generative, park2024generative}, but more work is  needed to understand how to best capture diverse user behaviors and develop ethical guidelines for deploying user simulations in safety evaluations \citep{agnew2024illusion, anthis2025llm}.

\subsection{How Can We Improve Researcher Access to Data for Understanding Interaction Harms?} Current public datasets of human-AI conversations, while valuable, fall short of meeting researchers' needs. Large collections of chat logs like WildChat and ShareGPT offer millions of conversations but provide only a narrow window into AI usage, capturing self-selected interactions that may not represent typical user behaviors or patterns \citep{zhao2024wildchat, ouyang2023shifted}. These datasets can be biased in terms of user demographics, use cases, and interaction styles, potentially leading to incomplete or incorrect conclusions about how people typically interact with AI systems. AI developers and service providers typically have access to the most comprehensive human-AI interaction data \citep{tamkin2024clio}. Given their incentives to protect user privacy and proprietary information, developing privacy-preserving data-sharing frameworks is essential to enable researchers to access representative interaction data while maintaining user trust and consent.

\subsection{What Infrastructure Do We Need to Facilitate Interactive Evaluations?} Many interactive evaluations share common stages: recruiting human participants, collecting interaction data, and administering surveys and other tasks. Developing accessible protocols, guides, and standardized test suites---similar to those available for static evaluations---can support and facilitate an increase in interactive evaluations \citep{UK_AI_Safety_Institute_2024, METR, collins2023evaluating}. Some existing platforms facilitate crowd-sourced interactive evaluations, but they may face validity concerns by gamifying safety testing (e.g., with tasks like 'break the model in one minute') and focusing primarily on vulnerabilities such as jailbreaking \citep{RedTeamArena2024}. Instead, we need more platforms that standardize the foundational elements of human subject studies with AI while maintaining scientific rigor, reducing technical overhead for researchers, and enabling broader participation in AI evaluation. Additionally, while current platforms primarily support single-session studies, understanding many interaction harms requires longitudinal evaluation capabilities with secure, privacy-preserving mechanisms for tracking behavioral changes over extended usage periods. This is a critical infrastructure gap that must be addressed to adequately assess the long-term influence of AI systems on human users.

\subsection{How Can Interactive Evaluations Produce Actionable Findings That Guide Stakeholder Decisions? What Are the Limitations of Controlled Studies in Capturing Broader Impacts?} Randomized control trials (RCTs) have proven valuable for policy decisions by providing rigorous evidence, particularly in public health and social policy \citep{hariton2018randomised}. Similarly, interactive evaluations aim to produce actionable findings about how AI systems influence human perception and behavior. These insights can help stakeholders make better decisions about model deployment, safety mechanisms, and interaction design. However, we currently lack adequate visibility into how different stakeholders---from AI labs to government bodies---actually use evaluation results in their decision-making processes. Bridging this gap requires closer collaboration between researchers and decision-makers to ensure evaluation designs align with practical needs \citep{hardy2024more}.

Finally, we must be mindful of limitations; like RCTs, controlled evaluations of human-AI interaction may effectively measure individual-level effects (like individual manipulation or overreliance) while still missing broader systemic patterns. Thus, while measuring these individual impacts is crucial for understanding immediate safety concerns, they should be complemented with examinations of how AI systems reshape institutional structures, professional practices, and social arrangements---effects that extend well beyond individual interactions \citep{Chater_Loewenstein_2023}.

\section{Conclusion}
As AI systems become increasingly integrated into our daily lives, examining harms that emerge through sustained engagement is essential for developing systems that prioritize human well-being. Our work makes three key contributions to this pursuit: first, we motivate the need to measure interaction harms and demonstrate why current evaluation paradigms systematically fail to capture them; second, we develop a structured framework of practical principles for designing interactive evaluations; and third, we identify specific implementation challenges and research directions that must be addressed for these methods to succeed at scale. Several established industries, from medicine to automotives, have long recognized the need to study their technologies' impact on people through extensive testing during development and after deployment \citep{10.1001/jama.2020.1166}. As AI capabilities and applications continue to expand, we must similarly strengthen our investment in understanding these systems' impact on human behavior and society at large. The methodological considerations outlined here provide a foundation for this shift in how we evaluate increasingly interactive AI systems.

\section*{Acknowledgments}
We are incredibly grateful to Seth Lazar, Katy Glenn Bass, and our anonymous peer reviewers for helpful comments on the paper. We also thank the participants of the Knight First Amendment Institute workshop on AI and democratic freedoms for their feedback. Finally, thank you to Jamie Bernardi, Merlin Stein, Patrick Levermore, Kobi Hackenburg, Deep Ganguli, Ben Bucknall, Esin Durmus, and Christopher Summerfield for helpful comments on an earlier draft of the paper.
\bibliography{aaai25}

@misc{UK_Government_2024, title={Ai Safety Institute approach to evaluations}, url={https://www.gov.uk/government/publications/ai-safety-institute-approach-to-evaluations/ai-safety-institute-approach-to-evaluations}, journal={GOV.UK}, author={{UK Government}}, year={2024}, month={Feb}}

@misc{OpenAI_2023, title={Preparedness}, url={https://openai.com/safety/preparedness}, journal={OpenAI}, author={OpenAI}, year={2023}, month={Dec}}

@misc{Anthropic_2023, title={Anthropic’s responsible scaling policy}, url={https://www.anthropic.com/news/anthropics-responsible-scaling-policy}, journal={Anthropic}, author={Anthropic}, year={2023}, month={Sep}}

@article{rottger2024safetyprompts,
  title={SafetyPrompts: a Systematic Review of Open Datasets for Evaluating and Improving Large Language Model Safety},
  author={R{\"o}ttger, Paul and Pernisi, Fabio and Vidgen, Bertie and Hovy, Dirk},
  journal={arXiv preprint arXiv:2404.05399},
  year={2024}
}

@article{chang2024survey,
  title={A survey on evaluation of large language models},
  author={Chang, Yupeng and Wang, Xu and Wang, Jindong and Wu, Yuan and Yang, Linyi and Zhu, Kaijie and Chen, Hao and Yi, Xiaoyuan and Wang, Cunxiang and Wang, Yidong and others},
  journal={ACM Transactions on Intelligent Systems and Technology},
  volume={15},
  number={3},
  pages={1--45},
  year={2024},
  publisher={ACM New York, NY}
}

@article{parrish2021bbq,
  title={BBQ: A hand-built bias benchmark for question answering},
  author={Parrish, Alicia and Chen, Angelica and Nangia, Nikita and Padmakumar, Vishakh and Phang, Jason and Thompson, Jana and Htut, Phu Mon and Bowman, Samuel R},
  journal={arXiv preprint arXiv:2110.08193},
  year={2021}
}

@article{weidinger2023sociotechnical,
  title={Sociotechnical safety evaluation of generative ai systems},
  author={Weidinger, Laura and Rauh, Maribeth and Marchal, Nahema and Manzini, Arianna and Hendricks, Lisa Anne and Mateos-Garcia, Juan and Bergman, Stevie and Kay, Jackie and Griffin, Conor and Bariach, Ben and others},
  journal={arXiv preprint arXiv:2310.11986},
  year={2023}
}

@article{liao2023rethinking,
  title={Rethinking model evaluation as narrowing the socio-technical gap},
  author={Liao, Q Vera and Xiao, Ziang},
  journal={arXiv preprint arXiv:2306.03100},
  year={2023}
}

@inproceedings{raji2022fallacy,
  title={The fallacy of AI functionality},
  author={Raji, Inioluwa Deborah and Kumar, I Elizabeth and Horowitz, Aaron and Selbst, Andrew},
  booktitle={Proceedings of the 2022 ACM Conference on Fairness, Accountability, and Transparency},
  pages={959--972},
  year={2022}
}

@article{ouyang2023shifted,
  title={The shifted and the overlooked: a task-oriented investigation of user-gpt interactions},
  author={Ouyang, Siru and Wang, Shuohang and Liu, Yang and Zhong, Ming and Jiao, Yizhu and Iter, Dan and Pryzant, Reid and Zhu, Chenguang and Ji, Heng and Han, Jiawei},
  journal={arXiv preprint arXiv:2310.12418},
  year={2023}
}

@inproceedings{weidinger2022taxonomy,
  title={Taxonomy of risks posed by language models},
  author={Weidinger, Laura and Uesato, Jonathan and Rauh, Maribeth and Griffin, Conor and Huang, Po-Sen and Mellor, John and Glaese, Amelia and Cheng, Myra and Balle, Borja and Kasirzadeh, Atoosa and others},
  booktitle={Proceedings of the 2022 ACM Conference on Fairness, Accountability, and Transparency},
  pages={214--229},
  year={2022}
}

@inproceedings{shelby2023sociotechnical,
  title={Sociotechnical harms of algorithmic systems: Scoping a taxonomy for harm reduction},
  author={Shelby, Renee and Rismani, Shalaleh and Henne, Kathryn and Moon, AJung and Rostamzadeh, Negar and Nicholas, Paul and Yilla-Akbari, N'Mah and Gallegos, Jess and Smart, Andrew and Garcia, Emilio and others},
  booktitle={Proceedings of the 2023 AAAI/ACM Conference on AI, Ethics, and Society},
  pages={723--741},
  year={2023}
}

@article{hartvigsen2022toxigen,
  title={Toxigen: A large-scale machine-generated dataset for adversarial and implicit hate speech detection},
  author={Hartvigsen, Thomas and Gabriel, Saadia and Palangi, Hamid and Sap, Maarten and Ray, Dipankar and Kamar, Ece},
  journal={arXiv preprint arXiv:2203.09509},
  year={2022}
}

@article{phuong2024evaluating,
  title={Evaluating Frontier Models for Dangerous Capabilities},
  author={Phuong, Mary and Aitchison, Matthew and Catt, Elliot and Cogan, Sarah and Kaskasoli, Alexandre and Krakovna, Victoria and Lindner, David and Rahtz, Matthew and Assael, Yannis and Hodkinson, Sarah and others},
  journal={arXiv preprint arXiv:2403.13793},
  year={2024}
}

@article{li2024wmdp,
  title={The wmdp benchmark: Measuring and reducing malicious use with unlearning},
  author={Li, Nathaniel and Pan, Alexander and Gopal, Anjali and Yue, Summer and Berrios, Daniel and Gatti, Alice and Li, Justin D and Dombrowski, Ann-Kathrin and Goel, Shashwat and Phan, Long and others},
  journal={arXiv preprint arXiv:2403.03218},
  year={2024}
}

@article{lee2022evaluating,
  title={Evaluating human-language model interaction},
  author={Lee, Mina and Srivastava, Megha and Hardy, Amelia and Thickstun, John and Durmus, Esin and Paranjape, Ashwin and Gerard-Ursin, Ines and Li, Xiang Lisa and Ladhak, Faisal and Rong, Frieda and others},
  journal={arXiv preprint arXiv:2212.09746},
  year={2022}
}

@article{collins2023evaluating,
  title={Evaluating language models for mathematics through interactions},
  author={Collins, Katherine M and Jiang, Albert Q and Frieder, Simon and Wong, Lionel and Zilka, Miri and Bhatt, Umang and Lukasiewicz, Thomas and Wu, Yuhuai and Tenenbaum, Joshua B and Hart, William and others},
  journal={arXiv preprint arXiv:2306.01694},
  year={2023}
}

@article{raji2021ai,
  title={AI and the everything in the whole wide world benchmark},
  author={Raji, Inioluwa Deborah and Bender, Emily M and Paullada, Amandalynne and Denton, Emily and Hanna, Alex},
  journal={arXiv preprint arXiv:2111.15366},
  year={2021}
}

@article{Chater_Loewenstein_2023, title={The i-frame and the s-frame: How focusing on individual-level solutions has led behavioral public policy astray}, volume={46}, DOI={10.1017/S0140525X22002023}, journal={Behavioral and Brain Sciences}, author={Chater, Nick and Loewenstein, George}, year={2023}, pages={e147}}

@incollection{2003xiii,
author = {Mike Kuniavsky},
booktitle = {Observing the User Experience},
publisher = {Morgan Kaufmann},
address = {San Francisco},
pages = {xiii-xvi},
year = {2003},
isbn = {978-1-55860-923-5},
doi = {https://doi.org/10.1016/B978-155860923-5/50027-9},
url = {https://www.sciencedirect.com/science/article/pii/B9781558609235500279}
}

@article{hariton2018randomised,
  author = {Hariton, Eduardo and Locascio, Joseph J.},
  title = {Randomised controlled trials - the gold standard for effectiveness research: Study design: randomised controlled trials},
  journal = {BJOG},
  year = {2018},
  volume = {125},
  number = {13},
  pages = {1716},
  doi = {10.1111/1471-0528.15199},
  pmid = {29916205},
  pmcid = {PMC6235704},
  month = {Dec},
  note = {Epub 2018 Jun 19}
}

@misc{zhao2024wildchat,
      title={WildChat: 1M ChatGPT Interaction Logs in the Wild}, 
      author={Wenting Zhao and Xiang Ren and Jack Hessel and Claire Cardie and Yejin Choi and Yuntian Deng},
      year={2024},
      eprint={2405.01470},
      archivePrefix={arXiv},
      primaryClass={cs.CL}
}

@misc{subramonyam2024bridging,
      title={Bridging the Gulf of Envisioning: Cognitive Design Challenges in LLM Interfaces}, 
      author={Hariharan Subramonyam and Roy Pea and Christopher Lawrence Pondoc and Maneesh Agrawala and Colleen Seifert},
      year={2024},
      eprint={2309.14459},
      archivePrefix={arXiv},
      primaryClass={cs.HC}
}

@misc{Mitchell_2024, title={Ethical ai isn’t to blame for Google’s Gemini debacle}, url={https://time.com/6836153/ethical-ai-google-gemini-debacle/}, journal={Time}, publisher={Time}, author={Mitchell, Margaret}, year={2024}, month={Feb}}

@misc{ibrahim2024characterizing,
      title={Characterizing and modeling harms from interactions with design patterns in AI interfaces}, 
      author={Lujain Ibrahim and Luc Rocher and Ana Valdivia},
      year={2024},
      eprint={2404.11370},
      archivePrefix={arXiv},
      primaryClass={cs.HC}
}

@inproceedings{10.1145/3531146.3533182,
author = {Liao, Q.Vera and Sundar, S. Shyam},
title = {Designing for Responsible Trust in AI Systems: A Communication Perspective},
year = {2022},
isbn = {9781450393522},
publisher = {Association for Computing Machinery},
address = {New York, NY, USA},
url = {https://doi.org/10.1145/3531146.3533182},
doi = {10.1145/3531146.3533182},
booktitle = {Proceedings of the 2022 ACM Conference on Fairness, Accountability, and Transparency},
pages = {1257–1268},
numpages = {12},
location = {<conf-loc>, <city>Seoul</city>, <country>Republic of Korea</country>, </conf-loc>},
series = {FAccT '22}
}

@misc{gao2024taxonomy,
  author = {Gao, J. and Gebreegziabher, S. A. and Choo, K. T. W. and Li, T. J. J. and Perrault, S. T. and Malone, T. W.},
  title = {A Taxonomy for Human-LLM Interaction Modes: An Initial Exploration},
  year = {2024},
  howpublished = {arXiv preprint arXiv:2404.00405}
}

@inproceedings{handler,
author = {Händler, Thorsten},
year = {2023},
month = {10},
pages = {},
title = {A Taxonomy for Autonomous LLM-Powered Multi-Agent Architectures},
doi = {10.5220/0012239100003598}
}

@ARTICLE{8404030,
  author={Damacharla, Praveen and Javaid, Ahmad Y. and Gallimore, Jennie J. and Devabhaktuni, Vijay K.},
  journal={IEEE Access}, 
  title={Common Metrics to Benchmark Human-Machine Teams (HMT): A Review}, 
  year={2018},
  volume={6},
  number={},
  pages={38637-38655},
  doi={10.1109/ACCESS.2018.2853560}}

@article{CORONADO2022392,
title = {Evaluating quality in human-robot interaction: A systematic search and classification of performance and human-centered factors, measures and metrics towards an industry 5.0},
journal = {Journal of Manufacturing Systems},
volume = {63},
pages = {392-410},
year = {2022},
issn = {0278-6125},
doi = {https://doi.org/10.1016/j.jmsy.2022.04.007},
url = {https://www.sciencedirect.com/science/article/pii/S0278612522000577},
author = {Enrique Coronado and Takuya Kiyokawa and Gustavo A. Garcia Ricardez and Ixchel G. Ramirez-Alpizar and Gentiane Venture and Natsuki Yamanobe}}

@inproceedings{10.1145/3411764.3445423,
author = {Gordon, Mitchell L. and Zhou, Kaitlyn and Patel, Kayur and Hashimoto, Tatsunori and Bernstein, Michael S.},
title = {The Disagreement Deconvolution: Bringing Machine Learning Performance Metrics In Line With Reality},
year = {2021},
isbn = {9781450380966},
publisher = {Association for Computing Machinery},
address = {New York, NY, USA},
url = {https://doi.org/10.1145/3411764.3445423},
doi = {10.1145/3411764.3445423},
booktitle = {Proceedings of the 2021 CHI Conference on Human Factors in Computing Systems},
articleno = {388},
numpages = {14},
location = {<conf-loc>, <city>Yokohama</city>, <country>Japan</country>, </conf-loc>},
series = {CHI '21}
}

@article{decisionregret,
author = {Brehaut, Jamie and O'Connor, Annette and Wood, Timothy and Hack, Tom and Siminoff, Laura and Gordon, Elisa and Feldman-Stewart, Deb},
year = {2003},
month = {08},
pages = {281-92},
title = {Validation of a Decision Regret Scale},
volume = {23},
journal = {Medical decision making : an international journal of the Society for Medical Decision Making},
doi = {10.1177/0272989X03256005}
}

@article{10.1001/jama.2020.1166,
    author = {Wouters, Olivier J. and McKee, Martin and Luyten, Jeroen},
    title = "{Estimated Research and Development Investment Needed to Bring a New Medicine to Market, 2009-2018}",
    journal = {JAMA},
    volume = {323},
    number = {9},
    pages = {844-853},
    year = {2020},
    month = {03},
    issn = {0098-7484},
    doi = {10.1001/jama.2020.1166},
    url = {https://doi.org/10.1001/jama.2020.1166},
    eprint = {https://jamanetwork.com/journals/jama/articlepdf/2762311/jama\_wouters\_2020\_oi\_200015\_1663357730.45628.pdf},
}

@misc{UK_AI_Safety_Institute_2024, title = {Inspect AI}, url={https://ukgovernmentbeis.github.io/inspect_ai/}, journal={Inspect}, author={{UK AI Safety Institute}}, year={2024}, month={May}}

@misc{METR, title ={METR Autonomy Evaluations Resources},  url={https://metr.github.io/autonomy-evals-guide/index.html}, journal={METR’s Autonomy Evaluation Resources}, author={{METR}}, year = {2023}}

@article{agnew2024illusion,
  title={The illusion of artificial inclusion},
  author={Agnew, William and Bergman, A Stevie and Chien, Jennifer and D{\'\i}az, Mark and El-Sayed, Seliem and Pittman, Jaylen and Mohamed, Shakir and McKee, Kevin R},
  journal={arXiv preprint arXiv:2401.08572},
  year={2024}
}

@misc{costello_pennycook_rand_2024,
 title={Durably reducing conspiracy beliefs through dialogues with AI},
 url={osf.io/preprints/psyarxiv/xcwdn},
 DOI={10.31234/osf.io/xcwdn},
 publisher={PsyArXiv},
 author={Costello, Thomas H and Pennycook, Gordon and Rand, David G},
 year={2024},
 month={Apr}
}

@inproceedings{Zhang_2024, series={CHI ’24},
   title={“It’s a Fair Game”, or Is It? Examining How Users Navigate Disclosure Risks and Benefits When Using LLM-Based Conversational Agents},
   url={http://dx.doi.org/10.1145/3613904.3642385},
   DOI={10.1145/3613904.3642385},
   booktitle={Proceedings of the CHI Conference on Human Factors in Computing Systems},
   publisher={ACM},
   author={Zhang, Zhiping and Jia, Michelle and Lee, Hao-Ping (Hank) and Yao, Bingsheng and Das, Sauvik and Lerner, Ada and Wang, Dakuo and Li, Tianshi},
   year={2024},
   month=may, collection={CHI ’24} }

@article{wallach2025position,
  title={Position: Evaluating Generative AI Systems is a Social Science Measurement Challenge},
  author={Wallach, Hanna and Desai, Meera and Cooper, A Feder and Wang, Angelina and Atalla, Chad and Barocas, Solon and Blodgett, Su Lin and Chouldechova, Alexandra and Corvi, Emily and Dow, P Alex and others},
  journal={arXiv preprint arXiv:2502.00561},
  year={2025}
}

@article{alberts2024should,
  title={Should agentic conversational AI change how we think about ethics? Characterising an interactional ethics centred on respect},
  author={Alberts, Lize and Keeling, Geoff and McCroskery, Amanda},
  journal={arXiv preprint arXiv:2401.09082},
  year={2024}
}

@article{mckee2024human,
  title={Human participants in AI research: Ethics and transparency in practice},
  author={McKee, Kevin R},
  journal={IEEE Transactions on Technology and Society},
  year={2024},
  publisher={IEEE}
}

@article{dev2021measures,
  title={On measures of biases and harms in NLP},
  author={Dev, Sunipa and Sheng, Emily and Zhao, Jieyu and Amstutz, Aubrie and Sun, Jiao and Hou, Yu and Sanseverino, Mattie and Kim, Jiin and Nishi, Akihiro and Peng, Nanyun and others},
  journal={arXiv preprint arXiv:2108.03362},
  year={2021}
}

@article{perez2022red,
  title={Red teaming language models with language models},
  author={Perez, Ethan and Huang, Saffron and Song, Francis and Cai, Trevor and Ring, Roman and Aslanides, John and Glaese, Amelia and McAleese, Nat and Irving, Geoffrey},
  journal={arXiv preprint arXiv:2202.03286},
  year={2022}
}

@inproceedings{blodgett2021stereotyping,
  title={Stereotyping Norwegian salmon: An inventory of pitfalls in fairness benchmark datasets},
  author={Blodgett, Su Lin and Lopez, Gilsinia and Olteanu, Alexandra and Sim, Robert and Wallach, Hanna},
  booktitle={Proceedings of the 59th Annual Meeting of the Association for Computational Linguistics and the 11th International Joint Conference on Natural Language Processing (Volume 1: Long Papers)},
  pages={1004--1015},
  year={2021}
}

@article{mccoy2023embers,
  title={Embers of autoregression: Understanding large language models through the problem they are trained to solve},
  author={McCoy, R Thomas and Yao, Shunyu and Friedman, Dan and Hardy, Matthew and Griffiths, Thomas L},
  journal={arXiv preprint arXiv:2309.13638},
  year={2023}
}

@article{ibrahim2025multi,
  title={Multi-turn Evaluation of Anthropomorphic Behaviours in Large Language Models},
  author={Ibrahim, Lujain and Akbulut, Canfer and Elasmar, Rasmi and Rastogi, Charvi and Kahng, Minsuk and Morris, Meredith Ringel and McKee, Kevin R and Rieser, Verena and Shanahan, Murray and Weidinger, Laura},
  journal={arXiv preprint arXiv:2502.07077},
  year={2025}
}

@article{hofmann2024ai,
  title={AI generates covertly racist decisions about people based on their dialect},
  author={Hofmann, Valentin and Kalluri, Pratyusha Ria and Jurafsky, Dan and King, Sharese},
  journal={Nature},
  volume={633},
  number={8028},
  pages={147--154},
  year={2024},
  publisher={Nature Publishing Group UK London}
}

@article{chen2024designing,
  title={Designing a dashboard for transparency and control of conversational AI},
  author={Chen, Yida and Wu, Aoyu and DePodesta, Trevor and Yeh, Catherine and Li, Kenneth and Marin, Nicholas Castillo and Patel, Oam and Riecke, Jan and Raval, Shivam and Seow, Olivia and others},
  journal={arXiv preprint arXiv:2406.07882},
  year={2024}
}

@article{staab2023beyond,
  title={Beyond memorization: Violating privacy via inference with large language models},
  author={Staab, Robin and Vero, Mark and Balunovi{\'c}, Mislav and Vechev, Martin},
  journal={arXiv preprint arXiv:2310.07298},
  year={2023}
}

@article{viegas2023system,
  title={The system model and the user model: Exploring AI dashboard design},
  author={Vi{\'e}gas, Fernanda and Wattenberg, Martin},
  journal={arXiv preprint arXiv:2305.02469},
  year={2023}
}

@article{ghandeharioun2025s,
  title={Who's asking? User personas and the mechanics of latent misalignment},
  author={Ghandeharioun, Asma and Yuan, Ann and Guerard, Marius and Reif, Emily and Lepori, Michael and Dixon, Lucas},
  journal={Advances in Neural Information Processing Systems},
  volume={37},
  pages={125967--126003},
  year={2025}
}

@article{jin2024implicit,
  title={Implicit personalization in language models: A systematic study},
  author={Jin, Zhijing and Heil, Nils and Liu, Jiarui and Dhuliawala, Shehzaad and Qi, Yahang and Sch{\"o}lkopf, Bernhard and Mihalcea, Rada and Sachan, Mrinmaya},
  journal={arXiv preprint arXiv:2405.14808},
  year={2024}
}

@article{reuel2024open,
  title={Open problems in technical ai governance},
  author={Reuel, Anka and Bucknall, Ben and Casper, Stephen and Fist, Tim and Soder, Lisa and Aarne, Onni and Hammond, Lewis and Ibrahim, Lujain and Chan, Alan and Wills, Peter and others},
  journal={arXiv preprint arXiv:2407.14981},
  year={2024}
}

@article{park2024generative,
  title={Generative agent simulations of 1,000 people},
  author={Park, Joon Sung and Zou, Carolyn Q and Shaw, Aaron and Hill, Benjamin Mako and Cai, Carrie and Morris, Meredith Ringel and Willer, Robb and Liang, Percy and Bernstein, Michael S},
  journal={arXiv preprint arXiv:2411.10109},
  year={2024}
}

@inproceedings{park2023generative,
  title={Generative agents: Interactive simulacra of human behavior},
  author={Park, Joon Sung and O'Brien, Joseph and Cai, Carrie Jun and Morris, Meredith Ringel and Liang, Percy and Bernstein, Michael S},
  booktitle={Proceedings of the 36th annual acm symposium on user interface software and technology},
  pages={1--22},
  year={2023}
}

@article{wang2025large,
  title={Large language models that replace human participants can harmfully misportray and flatten identity groups},
  author={Wang, Angelina and Morgenstern, Jamie and Dickerson, John P},
  journal={Nature Machine Intelligence},
  pages={1--12},
  year={2025},
  publisher={Nature Publishing Group UK London}
}

@article{RedTeamArena2024,
  title = {RedTeam Arena: An Open-Source, Community-driven Jailbreaking Platform},
  author = {Angelopoulos, Anastasios and Vivona, Luca and Chiang, Wei-Lin and Vichare, Aryan and Dunlap, Lisa and Salvivona and Pliny and Stoica, Ion},
  journal = {LMSYS Org Blog},
  year = {2024},
  month = {September},
  day = {13},
  url = {https://lmsys.org/blog/2024-09-13-redteam-arena/},
  note = {Accessed: 2025-03-05}
}

@article{hardy2024more,
  title={More than Marketing? On the Information Value of AI Benchmarks for Practitioners},
  author={Hardy, Amelia and Reuel, Anka and Meimandi, Kiana Jafari and Soder, Lisa and Griffith, Allie and Asmar, Dylan M and Koyejo, Sanmi and Bernstein, Michael S and Kochenderfer, Mykel J},
  journal={arXiv preprint arXiv:2412.05520},
  year={2024}
}

@article{phang2025investigating,
  title={Investigating Affective Use and Emotional Well-being on ChatGPT},
  author={Phang, Jason and Lampe, Michael and Ahmad, Lama and Agarwal, Sandhini and Fang, Cathy Mengying and Liu, Auren R and Danry, Valdemar and Lee, Eunhae and Chan, Samantha WT and Pataranutaporn, Pat and others},
  journal={arXiv preprint arXiv:2504.03888},
  year={2025}
}

@article{zhou2024haicosystem,
  title={Haicosystem: An ecosystem for sandboxing safety risks in human-ai interactions},
  author={Zhou, Xuhui and Kim, Hyunwoo and Brahman, Faeze and Jiang, Liwei and Zhu, Hao and Lu, Ximing and Xu, Frank and Lin, Bill Yuchen and Choi, Yejin and Mireshghallah, Niloofar and others},
  journal={arXiv preprint arXiv:2409.16427},
  year={2024}
}

@article{tamkin2024clio,
  title={Clio: Privacy-Preserving Insights into Real-World AI Use},
  author={Tamkin, Alex and McCain, Miles and Handa, Kunal and Durmus, Esin and Lovitt, Liane and Rathi, Ankur and Huang, Saffron and Mountfield, Alfred and Hong, Jerry and Ritchie, Stuart and others},
  journal={arXiv preprint arXiv:2412.13678},
  year={2024}
}

@article{anthis2025llm,
  title={LLM Social Simulations Are a Promising Research Method},
  author={Anthis, Jacy Reese and Liu, Ryan and Richardson, Sean M and Kozlowski, Austin C and Koch, Bernard and Evans, James and Brynjolfsson, Erik and Bernstein, Michael},
  journal={arXiv preprint arXiv:2504.02234},
  year={2025}
}

@misc{ftc_6b_companion_2025,
  author       = {{Federal Trade Commission}},
  title        = {FTC Launches Inquiry into {AI} Chatbots Acting as Companions},
  year         = {2025},
  howpublished = {Press release, U.S. Federal Trade Commission},
}

@misc{senate_judiciary_2025,
  author       = {{U.S. Senate Committee on the Judiciary}},
  title        = {Examining the Harm of {AI} Chatbots},
  year         = {2025},
  howpublished = {Hearing, Subcommittee on Crime and Counterterrorism, U.S. Senate},
}
\end{document}